\documentclass[a4paper,conference,final]{IEEEtran}
\usepackage{graphicx}
\usepackage{cite}
\usepackage{array}
\usepackage{amsmath}
\usepackage{amssymb}
\usepackage{textpos}
\usepackage{subfigure}
\usepackage{enumerate}
\usepackage{lscape}
\usepackage{float}
\usepackage{pdfsync}
\usepackage{epstopdf}
\usepackage{url}
\usepackage{color}
\usepackage{threeparttable}
\usepackage{multirow}
\usepackage{footnote}
\usepackage{tabu}

\begin{document}
\title{A Non-Stationary VVLC MIMO Channel Model for Street Corner Scenarios\\
~~\\
\vspace*{-7mm}\large{\emph{Invited Paper}}\vspace*{-3mm}}
%
%
%
\author{Qingshan Chen$^{1}$, Cheng-Xiang Wang$^{2,3,*}$, Jian Sun$^1$, Wensheng Zhang$^1$, and Qiuming Zhu$^4$\\
$^1$\small{Shandong Provincial Key Lab of Wireless Communication Technologies, School of Information Science and Engineering,}\\
\small{Shandong  University, Qingdao, Shandong, 266237, China.} \\
$^2$\small{National Mobile Communications Research Laboratory, School of Information Science and Engineering,} \\
\small{Southeast University, Nanjing, Jiangsu, 210096, China.}\\
$^3$\small{Purple Mountain Laboratories, Nanjing, 211111, China.} \\

$^4$\small{The Key Laboratory of Dynamic Cognitive System of Electromagnetic Spectrum Space,}\\
\small{College of Electronic and Information Engineering,}\\
\small{Nanjing University of Aeronautics and Astronautics, Nanjing 211106, China.}\\
\small{$^{*}$Corresponding author}\\
Email: chenqs1995@163.com, chxwang@seu.edu.cn, zhangwsh@sdu.edu.cn, sunjian@sdu.edu.cn, zhuqiuming@nuaa.edu.cn

}
\maketitle

\begin{abstract}
In recent years, the application potential of visible light communication (VLC) technology as an alternative and supplement to radio frequency (RF) technology has attracted people's attention. The study of the underlying VLC channel is the basis for designing the VLC communication system. In this paper, a new non-stationary geometric street corner model is proposed for vehicular VLC (VVLC) multiple-input multiple-output (MIMO) channel. The proposed model takes into account changes in vehicle speed and direction. The category of scatterers includes fixed scatterers and mobile scatterers (MS). Based on the proposed model, we derive the channel impulse response (CIR) and explore the statistical characteristics of the VVLC channel. The channel gain and root mean square (RMS) delay spread of the VVLC channel are studied. In addition, the influence of velocity change on the statistical characteristics of the model is also investigated. The proposed channel model can guide future vehicle-to-infrastructure (V2I) and vehicle-to-vehicle (V2V) optical communication system design.
\end{abstract}


\section{INTRODUCTION}
Every year the lives of approximately 1.35 million people are cuting short as a result of a road traffic crash. Road traffic injuries cause considerable economic losses to individuals, their families, and the whole country, which cost most countries 3$\%$ of their gross domestic product\cite{1}. Therefore, in order to reduce traffic accidents and economic losses, researchers pay more and more attention to the development of vehicle communication technologies. The vehicle network includes three main types of communications: infrastructure to vehicle (I2V), V2I, and V2V communications. The first two communication types to exchange information on external networks to obtain information about directions, routes, and traffic jams on each section. The V2V communication enables the sharing of information between vehicles, so as to adjust the driving state and avoid accidents.

In recent years, the advanced technology of vehicle ad-hoc network (VANET) can improve road safety and traffic efficiency. To implement dedicated short range communications/wireless access in vehicular environments standards, the addition of new hardware is essential. Nevertheless, this will increase production costs. Since VLC system does not require too much additional hardware, it can be used as an alternative to RF communication between vehicles. The optical signals transmitters (Txs) the portion of VLC systems typically uses commercially available off-the-shelf incoherent light-emitting diodes (LEDs). The optical signal receivers (Rxs) portion of VLC systems typically uses high sensitive photodiodes (PDs) or camera-based receivers. In street corner scenarios, road visibility is generally poor for drivers. Real-times road condition information can be provided by V2V communication technology with extremely low latency to help avoid traffic collisions. A detailed and accurate understanding of the channel model of the  street scenarios is helpful for system design and performance assessment.


%
A comprehensive and detailed overview of the existing visible light communication channel model can be found in \cite{16,18}. Existing VLC channel models include indoor scenarios channel models, outdoor scenarios channel models, and underwater scenarios channel models. For indoor scenarios, the ray tracing method can well establish a deterministic model in a specific indoor environment to obtain the received optical power, delay of direct/indirect light and CIR \cite{2,3,4}. Deterministic channels modeling to require a time-consuming and specific characterization of the propagation condition, so it can not be easily extended to a broader range of scenarios. The method of geometric modeling can well obtain the statistical characteristics of channels\cite{5,6}. In \cite{5}, a geometry-based single bounce model for VLC channels was proposed. In \cite{6}, a regular-shaped geometry-based multiple bounce model (RS-GBMB) channel model is described, and the statistical characteristics of the line-of-sight (LoS), single-, double-, and triple bounced rays are compared. M. Elamassie et al. develop a path loss expression with water type, beam divergence angle, and receiver aperture diameter as parameters to determine the maximum achievable distance of different types of water\cite{7}. In\cite{8}, the ray tracing technique is used to develop a path loss model of the V2V link under different weather conditions, and the maximum achievable distance can be determined while ensuring a given bit error rate. The non-stationary RS-GBSM model proposed in \cite{9} proves that the optical power carried by the double-bounce (DB) components is too low and can be ignored. However, models in \cite{8} and \cite{9} are single-input single-output (SISO) channel models. A three-dimensional non-stationary channel model suitable for the fifth generation (5G) V2V communication scenario is proposed in \cite{17}. In \cite{10}, the modular structure is used for channel modeling and  $2 \times 2$ MIMO is compared to other MIMO configurations. An RF-based street corner channel model was proposed in \cite{11}. However, there is no accurate street corner channel model to describe the characteristics of VLC.


In this paper, we use geometry-based methods to derive a new non-stationary MIMO channel model for street corner scenarios. The model considers moving scatterers and fixed scatterers, corresponding to moving pedestrians (vehicles) and roadside trees (buildings), respectively. Angle of departures (AoDs) and angle of arrivals (AoAs) are time-varying due to the movement of the vehicle and the moving scatterers, resulting in a non-stationary model.

The main contributions of this paper are as follows:

(1) A new non-stationary MIMO GBSM for VVLC channels suitable for street corner scenes is proposed. At the same time, the speed of the vehicle and the changes of the scatterer are taken into account, making the model more versatile and practical.

(2) Using the proposed model, we obtained the statistical properties of the VVLC channel, i.e., path lengths experienced by light, CIR, channel gain, and RMS delay spread.

The remainder of the paper is structured as follows. Section II gives a brief introduction about the geometric channel model of the street corner. The VVLC channel coefficient is derived in Section III. Section IV presents the simulation results and analysis. Finally, the main conclusions of the paper are summarized in Section V.

\section{A VVLC CHANNEL MODEL FOR STREET CORNER SCENARIOS}

Road traffic, road environment, and wave propagation between vehicles are the three major elements of VVLC channel model. A typical street corner road scene is shown in Fig.~1. Road environment includes roadside environment such as buildings, road signs, and trees, as well as dynamic road traffic such as moving pedestrians and vehicles. Tx mode and Rx aperture size are two key elements that need to be a model of wave propagation. However, owing to VVLC technology is none the less in the initial phases of the study, there is no normative headlamp with a measuring beam pattern model. Therefore, we assume that each LED lamp in this paper conforms to the ideal Lambertian radiation pattern, and the radiation intensity can be expressed as\cite{12}

\setlength{\unitlength}{1mm}\label{1}
\begin{center}\begin{figure}[h]
\scalebox{-0.27}{\rotatebox{180}{\includegraphics{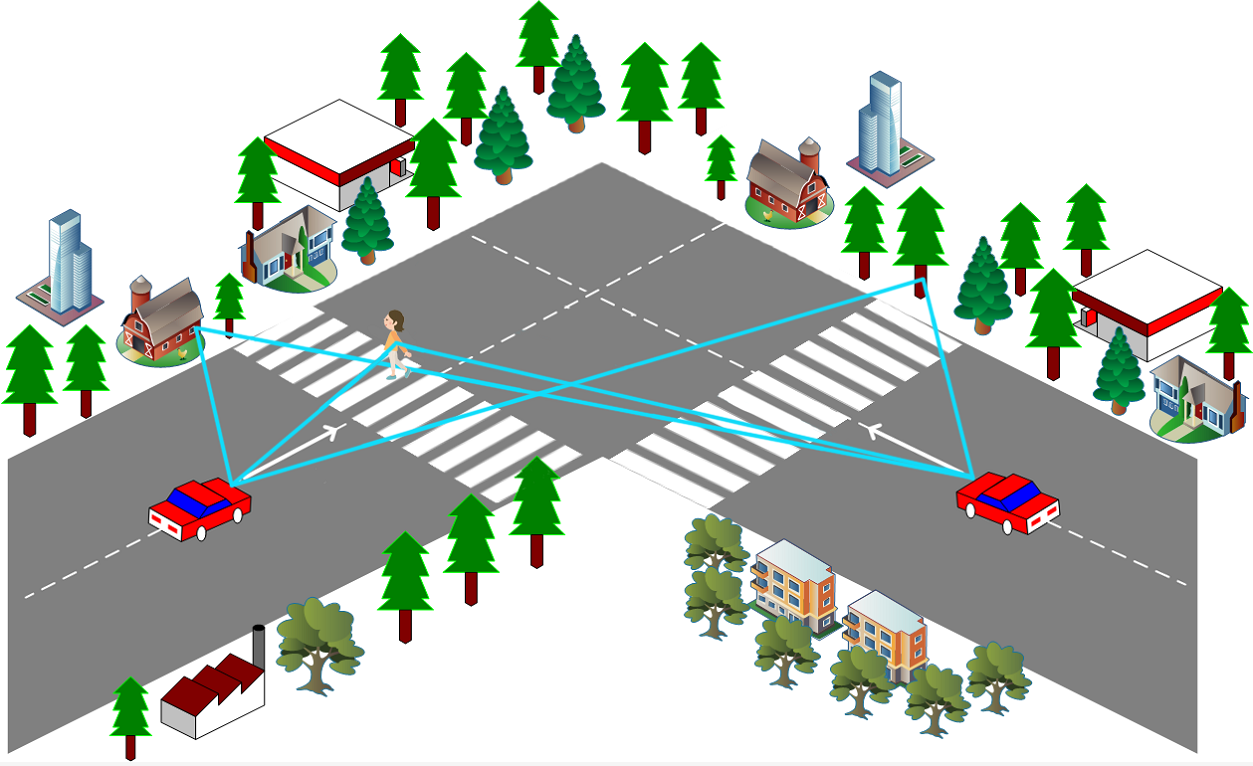}}}
\caption{The typical street corner road propagation environment.}
\end{figure}\end{center}
\vspace*{-13mm}

\setlength{\unitlength}{1mm}\label{2}
\begin{center}\begin{figure}[h]
\scalebox{-0.75}{\rotatebox{180}{\includegraphics{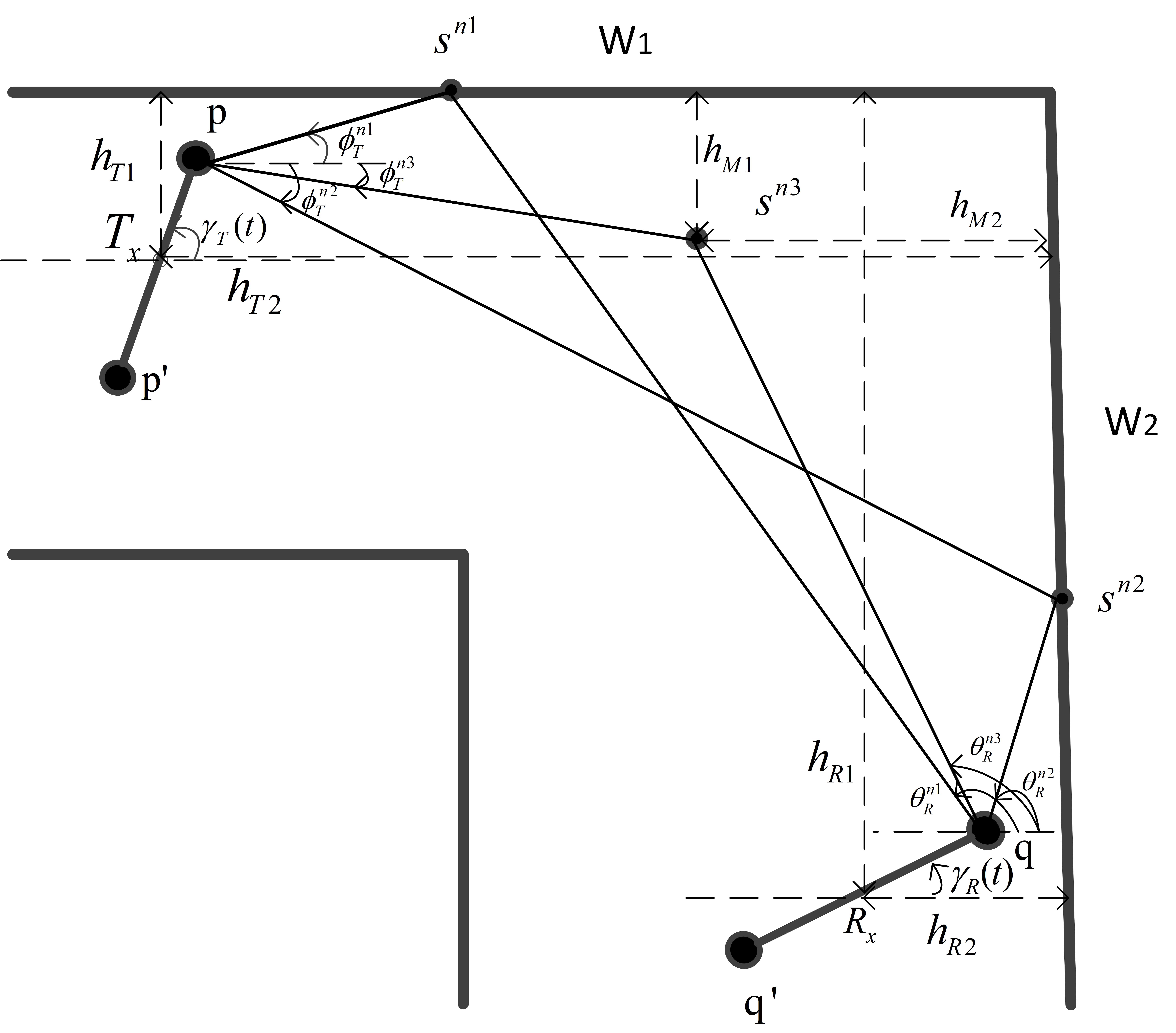}}}
\caption{The geometric corner channel model includes moving and fixed scatterers.}
\end{figure}\end{center}

\begin{equation}\label{1}
 R\left(\phi_{\mathrm{T}}\right)=\frac{\alpha+1}{2 \pi} \cos ^{\alpha}\left(\phi_{\mathrm{T}}\right), \quad \phi_{\mathrm{T}} \in[-\pi / 2, \pi / 2]
\end{equation}
here, $\alpha$ is the mode number of the radiation lobe, which specifies the directivity of the light source, and $\phi_{\mathrm{T}}$ is the angle of irradiance which is generally denoted as the AoD. At the receiver, assuming that the optical receiver is a non-imaging p-type/intrinsic/n-type (PIN) PD equipped at the vehicle.

\begin{table*}[htbp]\label{2}
\caption{DEFINITION OF SIGNIFICANT PARAMETERS IN THE MODEL.}
\begin{center}
\begin{tabular}{|c|c|}
\hline
${\delta _T},{\delta _R}$ & spacing between antenna elements at Tx and Rx\\
\hline
${\gamma _T},{\gamma _R}$ & tilt angle of the Tx and Rx, respectively\\
\hline
${v _T},{v _R},{v _M}$ & velocity of Tx , Rx and MS, respectively\\
\hline
$\phi_{\mathrm{T}}^{\left(n_{i}\right)}(i=1,2,3)$ & AoD of the waves that emitted from the Tx\\
\hline
$\theta_{\mathrm{R}}^{\left(n_{i}\right)}(i=1,2,3)$ & AoA of the waves from effective scatterers $S^{\left(n_{i}\right)}$ to the Rx\\
\hline
$\phi_{\mathrm{S}}^{\left(n_{i}\right)}(i=1,2,3)$ & AoD of the waves from the effective scatterers $S^{\left(n_{i}\right)}$\\
\hline
$\theta_{\mathrm{S}}^{\left(n_{i}\right)}(i=1,2,3)$ & AoA of the waves from the Tx to the effective scatterers $S^{\left(n_{i}\right)}$\\
\hline
$\varepsilon_{\mathrm{Tx}-\mathrm{S}^{n} i}$ & distances from the scatterers to Tx $S^{\left(n_{i}\right)},(i=1,2,3)$\\
\hline
$\varepsilon_{\mathrm{S}^{n} i-\mathrm{Rx}}$ & distances from Rx to scatterers $S^{\left(n_{i}\right)},(i=1,2,3)$\\
\hline
$h_{T_{i}}$ & distances from the $W_{i}$ $(\mathrm{i}=1,2)$ to Tx\\
\hline
$h_{R_{i}}$ & distances from the $W_{i}$ $(\mathrm{i}=1,2)$ to Rx\\
\hline
$h_{M_{i}}$ & distances from the $W_{i}$ $(\mathrm{i}=1,2)$ to MS\\
\hline
\end{tabular}
\label{tab1}
\end{center}
\end{table*}

\newcommand{\tabincell}[2]{\begin{tabular}{@{}#1@{}}#2\end{tabular}}
\begin{table}\label{1}
\caption{POSSIBLE OPTICAL PATHS.}
\centering
\begin{tabular}{|c|c|c|}
\hline
\textbf{Component} & \textbf{Experience Path} & \textbf{Length}\\
\hline
$\mathrm{SB} -11$& $\mathrm{Tx}_{\mathrm{p} / \mathrm{q}} \rightarrow S_{n_{1}} \rightarrow \mathrm{Rx}_{\mathrm{p} / \mathrm{q}}$ & \tabincell{c}{$\varepsilon_{\mathrm{p}-n_{1}}+\varepsilon_{n_{1}-\mathrm{q}}$ ,\\
$\varepsilon_{\mathrm{p}^{\prime}-n_{1}}+\varepsilon_{n_{1}-\mathrm{q}}$}\\
\hline
$\mathrm{SB} -12$ & $\mathrm{Tx}_{\mathrm{p} / \mathrm{q}} \rightarrow S_{n_{2}} \rightarrow \mathrm{Rx}_{\mathrm{p} / \mathrm{q}}$ & \tabincell{c}{$\varepsilon_{\mathrm{p}-n_{2}}+\varepsilon_{n_{2}-\mathrm{q}}$ ,\\ $\varepsilon_{\mathrm{p}^{\prime}-n_{2}}+\varepsilon_{n_{2}-\mathrm{q}}$}\\
\hline
$\mathrm{SB} -13$ & $\mathrm{Tx}_{\mathrm{p} / \mathrm{q}} \rightarrow S_{n_{3}} \rightarrow \mathrm{Rx}_{\mathrm{p} / \mathrm{q}}$ & \tabincell{c}{$\varepsilon_{\mathrm{p}-n_{3}}+\varepsilon_{n_{3}-\mathrm{q}}$ ,\\
$\varepsilon_{\mathrm{p}^{\prime}-n_{3}}+\varepsilon_{n_{3}-\mathrm{q}}$}\\
\hline
\end{tabular}
\end{table}

It has an effective acquisition area called the field of view (FoV), the rays can only be captured when they are radiated inside the receiver¡¯s FoV. The effective collection area $A_{\mathrm{R}_{\mathrm{eff}}}$ is given as\cite{12}

\begin{equation}\label{2}
A_{\mathrm{R}_{\mathrm{eff}}}=\left\{\begin{array}{cc}{A_{R} \cos \left(\theta_{\mathrm{R}}\right)} & {0 \leq \theta_{\mathrm{R}} \leq \mathrm{FoV}} \\ {0} & {\theta_{\mathrm{R}}>\mathrm{FoV}}\end{array}\right.
\end{equation}
here, ${A_{R}}$ is the area of the PD. The proposed VVLC street corner channel model is shown in Fig. 2. It is assumed that there are $N_{1}$ effective scatterers distributed on the buildings $W_{1}$ at the top side of this figure, which are denoted by $S^{\left(n_{1}\right)}\left(n_{1}=1, \dots, N_{1}\right)$, there are $N_{2}$ effective scatterers distributed on the buildings $W_{2}$  at the right side of Fig. 2, which are denoted by $S^{\left(n_{2}\right)}\left(n_{2}=1, \dots, N_{2}\right)$. Pedestrians and vehicles moving in the middle of the road are modeled as $N_{3}$ effective scatterers which are denoted by $S^{\left(n_{3}\right)}\left(n_{3}=1, \dots, N_{3}\right)$. $S^{\left(n_{3}\right)}$ can move with a time-varying velocities including a time-varying speed and a time-varying move angle. Tx is moving in the directions ${\gamma _T}$ with the speed ${v _T}$, and similarly, Rx is moving in the directions ${\gamma _R}$ with the speed ${v _R}$. At the Tx, $\phi_{\mathrm{T}}^{n_{1}}$, $\phi_{\mathrm{T}}^{n_{2}}$, and $\phi_{\mathrm{T}}^{n_{3}}$ are the AoDs. The $\theta_{\mathrm{S}}^{n_{i}}$ and $\phi_{\mathrm{S}}^{n_{i}}$ represent the AoAs and AoDs that from the effective scatterers side $S^{\left(n_{i}\right)}$. On the other hand, $\theta_{\mathrm{R}}^{n_{1}}$, $\theta_{\mathrm{R}}^{n_{2}}$, and $\theta_{\mathrm{R}}^{n_{3}}$ represent AoAs of the waves  that arrive at the Rx. In this paper, it is assumed that the AoDs $\phi_{\mathrm{T}}^{n_{i}}$ of isotropic scatterers around $W_{1}$ and $W_{2}$ follow a von Mises distribution, and only single bounce (SB) components propagation conditions are taken into account, since the optical power carried by the DB component can be ignored \cite{11}. The definitions of significant geometry parameters and the Possible optical paths are detailed in Table~I and Table~II, respectively.


\section{ VVLC CHANNEL COEFFICIENT }

The proposed channel model expresses CIR between the Tx and the Rx as the superposition of the non-LoS (NLoS) components, which could be described as

\begin{equation}\label{3}
h_{p q}(t)=h_{p q}^{\mathrm{NLoS}}(t)=\sum_{i=1}^{I} h_{p q}^{\mathrm{SB}_{i}}(t).
\end{equation}

 Here, $I=3$ means that SB rays have three subcomponents, i.e., SB-11 scattered by the $W_{1}$, SB-12 scattered by the $W_{2}$, and SB-13 scattered by the moving vehicles or pedestrians. In order to build a more accurate VVLC channel model, we introduce the $V\left(\Psi\right)$ visible function that is only received by scatterers distributed within PD's FoV at the Rx side. The visible function can be expressed as

\begin{equation}\label{4}
V\left(\theta_{\mathrm{R}}\right)=\left\{\begin{array}{ll}{1} & {\text { if } 0 \leq \theta_{\mathrm{R}} \leq \text { FoV }} \\ {0} & {\text { if } \theta_{\mathrm{R}}>\text { FoV }}\end{array}\right..
\end{equation}

\subsection{Single-Bounce Channel caused by walls}

The first possible light path in the geometric corner model is SB-11. The CIR of the link SB-11 can be written as\cite{6}

\begin{equation}\label{5}
\begin{split}
h_{pq}^{{\rm{S}}{{\rm{B}}_1}}(t) = &\sum\limits_{{n_1} = 1}^{{N_1}} {\frac{{(\alpha  + 1)V\left( {{\theta_{\mathrm{R}}}} \right)}}{{2\pi {{\left( {{\varepsilon _p}_{ - {{\rm{S}}^{{n_1}}}}} \right)}^2}}}} {\cos ^\alpha }\left( {\phi _{\rm{T}}^{{n_1}}} \right)\cos \left( {\theta _{\rm{S}}^{{n_1}}} \right)\\
& \times \frac{{(\alpha  + 1){A_{\rm{R}}}\rho_{\mathrm{Wall}} }}{{2\pi {{\left( {{\varepsilon _{{{\rm{S}}^{{n_1}}}}}_{ - q}} \right)}^2}}}{\cos ^\alpha }\left( {\phi _{\rm{S}}^{{n_1}}} \right)\cos \left( {\theta _{\rm{R}}^{{n_1}}} \right)\\
& \times \delta \left( {t - \frac{{{\varepsilon _p}_{ - {{\rm{S}}^{{n_1}}}} + {\varepsilon _{{{\rm{S}}^{{n_1}}}}}_{ - q}}}{c}} \right).
\end{split}
\end{equation}

Similarly, the CIR of the other possible light path SB-12 can be written as

\begin{equation}\label{6}
\begin{split}
h_{pq}^{{\rm{S}}{{\rm{B}}_2}}(t) = &\sum\limits_{{n_2} = 1}^{{N_2}} {\frac{{(\alpha  + 1)V\left( {{\theta_{\mathrm{R}}}} \right)}}{{2\pi {{\left( {{\varepsilon _p}_{ - {{\rm{S}}^{{n_2}}}}} \right)}^2}}}} {\cos ^\alpha }\left( {\phi _{\rm{T}}^{{n_2}}} \right)\cos \left( {\theta _{\rm{S}}^{{n_2}}} \right)\\
& \times \frac{{(\alpha  + 1){A_{\rm{R}}}\rho_{\mathrm{Wall}} }}{{2\pi {{\left( {{\varepsilon _{{{\rm{S}}^{{n_2}}}}}_{ - q}} \right)}^2}}}{\cos ^\alpha }\left( {\phi _{\rm{S}}^{{n_2}}} \right)\cos \left( {\theta _{\rm{R}}^{{n_2}}} \right)\\
& \times \delta \left( {t - \frac{{{\varepsilon _p}_{ - {{\rm{S}}^{{n_2}}}} + {\varepsilon _{{{\rm{S}}^{{n_2}}}}}_{ - q}}}{c}} \right).
\end{split}
\end{equation}

Here, $\rho_{\mathrm{Wall}}$ is the reflection coefficient of the wall, $\delta(.)$ refers to the Dirac delta function, $c$ is the speed of light. From Figure 2, according to the application of the cosine theorem to an appropriate triangle, we get

\begin{equation}\label{7}
{\varepsilon _{T - {{\rm{S}}^{{n_1}}}}} = \frac{{{h_{T1}}}}{{\sin \phi _T^{{n_1}}}}
\end{equation}

\begin{equation}\label{8}
{\varepsilon _{p - {{\rm{S}}^{{n_1}}}}} = {\varepsilon _{T - {{\rm{S}}^{{n_1}}}}} - \frac{1}{2}{\delta _T}\cos \left( {{\gamma _T} - \phi _T^{{n_1}}} \right)
\end{equation}

\begin{equation}\label{9}
{\varepsilon _{R - {{\rm{S}}^{{n_1}}}}} = \frac{{{h_{R1}}}}{{\sin \theta _R^{{n_1}}}}
\end{equation}

\begin{equation}\label{10}
{{\varepsilon _{{{\rm{S}}^{{n_1}}}}}_{ - q}} = {\varepsilon _{R - {{\rm{S}}^{{n_1}}}}} - \frac{1}{2}{\delta _R}\cos \left( {\theta _R^{{n_1}}-{\gamma _R}} \right)
\end{equation}

\begin{equation}\label{11}
{\varepsilon _{T - {{\rm{S}}^{{n_2}}}}} = \frac{{{h_{T2}}}}{{\cos \phi _T^{{n_1}}}}
\end{equation}

\begin{equation}\label{12}
{\varepsilon _{p - {{\rm{S}}^{{n_2}}}}} = {\varepsilon _{T - {{\rm{S}}^{{n_2}}}}} - \frac{1}{2}{\delta _T}\cos \left( {{\gamma _T} - \phi _T^{{n_2}}} \right)
\end{equation}

\begin{equation}\label{13}
{\varepsilon _{R - {{\rm{S}}^{{n_2}}}}} = \frac{{{h_{R2}}}}{{\cos \theta _R^{{n_2}}}}
\end{equation}

\begin{equation}\label{14}
{{\varepsilon _{{{\rm{S}}^{{n_2}}}}}_{ - q}} = {\varepsilon _{R - {{\rm{S}}^{{n_2}}}}} - \frac{1}{2}{\delta _R}\cos \left( {\theta _R^{{n_2}}-{\gamma _R}} \right).
\end{equation}

In the proposed geometric model, $\phi_{T}^{\left(n_{1}\right)}$ is in the range of $(0,\pi / 2)$, and $\phi_{T}^{\left(n_{2}\right)}$ is in the range of $(-\pi / 2,0)$. AoDs and AoAs are correlated. For fixed clusters, we can get the relationship between $\phi_{T}^{\left(n_{1,2}\right)}$ and $\theta_{T}^{\left(n_{1,2}\right)}$

\begin{equation}\label{15}
\theta _R^{{n_1}} = \pi  - \arctan \frac{{{h_{R1}}}}{{{h_{T2}} - \frac{{{h_{T1}}}}{{\tan \phi _T^{{n_1}}}} - {h_{R2}}}}
\end{equation}

\begin{equation}\label{16}
\theta _R^{{n_2}} = \arctan \frac{{{h_{R1}} - {h_{T1}} + {h_{T2}} \cdot \tan \phi _T^{{n_2}}}}{{{h_{R2}}}}.
\end{equation}

\subsection{Single-Bounce Channel caused by  moving vehicles or pedestrians }

According to our proposed model, moving vehicles or pedestrians can create new possible light paths, i.e., link SB-13. The CIR for the light path of SB-13 can be written as

\begin{equation}\label{17}
\begin{split}
h_{pq}^{{\rm{S}}{{\rm{B}}_3}}(t) = &\sum\limits_{{n_3} = 1}^{{N_3}} {\frac{{(\alpha  + 1)V\left( {{\theta_{\mathrm{R}}}} \right)}}{{2\pi {{\left( {{\varepsilon _p}_{ - {{\rm{S}}^{{n_3}}}}} \right)}^2}}}} {\cos ^\alpha }\left( {\phi _{\rm{T}}^{{n_3}}} \right)\cos \left( {\theta _{\rm{S}}^{{n_3}}} \right)\\
& \times \frac{{(\alpha  + 1){A_{\rm{R}}}\rho_{\mathrm{Vehicles}} }}{{2\pi {{\left( {{\varepsilon _{{{\rm{S}}^{{n_3}}}}}_{ - q}} \right)}^2}}}{\cos ^\alpha }\left( {\phi _{\rm{S}}^{{n_3}}} \right)\cos \left( {\theta _{\rm{R}}^{{n_3}}} \right)\\
& \times \delta \left( {t - \frac{{{\varepsilon _p}_{ - {{\rm{S}}^{{n_3}}}} + {\varepsilon _{{{\rm{S}}^{{n_3}}}}}_{ - q}}}{c}} \right).
\end{split}
\end{equation}

The reflection coefficient of the vehicles body is expressed by $\rho_{\mathrm{Vehicles}}$. The optical path length can be expressed by the law of cosines in triangle $\mathrm{T} \mathrm{x}-\mathrm{S}^{n_{3}}-\mathrm{R} \mathrm{x}$

\begin{equation}\label{18}
{\varepsilon _{T - {{\rm{S}}^{{n_3}}}}} = \sqrt {{{\left[ {{h_{M1}} - {h_{T1}}} \right]}^2} + {{\left[ {{h_{M2}} - {h_{T2}}} \right]}^2}}
\end{equation}

\begin{equation}\label{19}
{\varepsilon _{p - {{\rm{S}}^{{n_3}}}}} = {\varepsilon _{T - {{\rm{S}}^{{n_3}}}}} - \frac{1}{2}{\delta _T}\cos \left( {{\gamma _T} - \phi _T^{{n_3}}} \right)
\end{equation}

\begin{equation}\label{20}
{\varepsilon _{R - {{\rm{S}}^{{n_3}}}}} = \sqrt {{{\left[ {{h_{M1}} - {h_{R1}}} \right]}^2} + {{\left[ {{h_{M2}} - {h_{R2}}} \right]}^2}}
\end{equation}

\begin{equation}\label{21}
{{\varepsilon _{{{\rm{S}}^{{n_3}}}}}_{ - q}} = {\varepsilon _{R - {{\rm{S}}^{{n_3}}}}} - \frac{1}{2}{\delta _R}\cos \left( {\theta _R^{{n_3}}-{\gamma _R}} \right).
\end{equation}

The moving cluster $S^{\left(n_{3}\right)}$ can be thought of as a disc with a certain angular distribution on a horizontal plane. The relationship between $\phi_{T}^{\left(n_{3}\right)}$ and $\theta_{T}^{\left(n_{3}\right)}$ can be determined as

\begin{equation}\label{22}
\theta _R^{{n_3}} = \arctan \left( {\frac{{\tan \left( {\phi _T^{{n_3}} - \mu _T^{{n_3}}} \right) \cdot {\varepsilon _{T - {{\rm{S}}^{{n_3}}}}}}}{\varepsilon _{R - {{\rm{S}}^{{n_3}}}}}} \right) + \mu _R^{{n_3}},
\end{equation}

where

\begin{equation}\label{23}
\mu _T^{{n_3}} = \arctan \left( {\frac{{{h_{M1}} - {h_{T1}}}}{{{h_{T2}} - {h_{M2}}}}} \right)
\end{equation}

\begin{equation}\label{24}
\mu _R^{{n_3}} = \arctan \left( {\frac{{{h_{M2}} - {h_{R2}}}}{{{h_{R1}} - {h_{M1}}}}} \right).
\end{equation}

\section{ SIMULATION RESULTS AND ANALYSES }

In this section, we demonstrate the simulation of the proposed geometric corner scattering model, and analyze the channel properties, including CIR, zero-frequency (DC) gain, and RMS delay spread. For simulation, the values of the parameters are set as follow, $\delta_{T}=\delta_{R}=0.5$ m, ${v _T}={v _R}=7$ m/s, ${v _M}=1$ m/s, $h_{T_{1}}=h_{R_{2}}=40$ m, $h_{R_{1}}=h_{T_{2}}=3$ m, $h_{M_{1}}=h_{M_{2}}=8$ m, $\rho_{\mathrm{Vehicles}}= 0.8$, $\rho_{\mathrm{Wall}}= 0.4$, $FOV=80^{\circ}$\cite{9}, $A_{\mathrm{R}}= 1 \mathrm{cm}^{2}$, ${\gamma _T}=0$, ${\gamma _R}=\pi/2$, $\alpha=1$, $c= 3\times \mathrm{10}^{8}$ m.

\subsection{VVLC CIR}

The resulting VVLC CIR of $h_{p q}^{\mathrm{SB}_{1}}(t)$, $h_{p q}^{\mathrm{SB}_{2}}(t)$, and $h_{p q}^{\mathrm{SB}_{3}}(t)$ components are shown in Fig. 3. We can quickly notice that the CIR of the SB-13 component is lower than that of SB-11 and SB-12. The reason is that the travel distance of SB-13 is longer than that of SB-11 and SB-12, so the path loss of light is also larger, which is why SB-13 has a larger time delay than the other two. However, roadside buildings have lower reflectivity and absorb more incident energy. Therefore, the increase in the optical path length is the reason for the reduction of signal strength. Fig. 4 is the total CIR of the SB component of the geometric street corner model. It is the sum of SB-11, SB-12 and SB-13.

The discussion above was based on the assumption of static conditions, i.e., t = 0. We assume the dynamic conditions that Tx and Rx maintain the original direction of movement and the same speed of change. For the geometric corner channel model, Fig. 5 shows the variation of CIR with vehicle motion. As the distance between the Tx vehicle and the target Rx vehicle continues to decrease, the received power gradually increases and the time delay decreases. Since the speed of the vehicle will gradually slow down during actual driving, the change of the received signal will also slow down accordingly.

\subsection{DC Channel Gain of VVLC}

For VVLC, channel DC gain is an important channel feature, which is completely characterized by CIR. The zero frequency value of the frequency response of the intensity-input channel and the intensity-output channel can be expressed as\cite{9}

\begin{equation}\label{25}
H(0)=\int_{-\infty}^{\infty} h(\tau) d \tau.
\end{equation}

The channel DC gain is part of the power emitted by the continuous wave transmitter detected at the receiving end. The channel gain in the logarithmic domain can be described as\cite{13}

\begin{equation}\label{26}
\text { Optical gain }(\mathrm{dB})=-10 \log _{10} H(0).
\end{equation}

We evaluate the mean DC channel gain and the channel gain distribution at a specific length, in dB, with the specific parameters set as above. Based on the simulated data, we next analyze the propagation characteristics of large collection scatterer locations with different angular distributions. Fig. 6 is the channel gain distribution of the SB component synthesized by the geometric street corner model. It can be clearly seen that the channel fading is larger compared to the result of Ahmed's model and the mean channel gain is -94.7dB. The reason is that the street corner scenario has a longer transmission distance. At the same time, we can see that the channel DC gain conforms to the gaussian distribution\cite{12}.

\subsection{RMS Delay Spread}

Under the influence of multipath, the actual received signal is the sum of the weighted and delayed transmission signals. Therefore, the time dispersion of the VVLC channel is caused by the time stretching of the transmitted signal. RMS delay spread of the channel can quantify time dispersion\cite{6}

\begin{equation}\label{27}
D_{\mathrm{rms}}=\sqrt{\frac{\int_{-\infty}^{\infty}\left(t-\mu_{\tau}\right)^{2} h^{2}(t) d t}{\int_{-\infty}^{\infty} h^{2}(t) d t}},
\end{equation}
where $\mu_{\tau}$ is the mean excess delay

\begin{equation}\label{28}
\mu_{\tau}=\frac{\int_{-\infty}^{\infty} t h^{2}(t) d t}{\int_{-\infty}^{\infty} h^{2}(t) d t}.
\end{equation}

The bandwidth of the VVLC channel is directly determined by the time dispersion, where the maximum bit rate $R_{b} \leq 1 / 10 D_{\mathrm{rms}}$\cite{14}. Therefore, in order to avoid intersymbol interference (ISI), it is usually necessary to limit the symbol length\cite{15}. The RMS delay spread of VVLC channel is shown in Fig. 7. We can see that the RMS delay spread of the street model is about 30 times that of the Ahmed's model, because in the street model, the SB-13 component a larger proportion and all the SB paths are longer. Similarly, they all obey the Gaussian distribution.

\setlength{\unitlength}{1mm}\label{3}
\begin{center}\begin{figure}[h]
\scalebox{-0.59}{\rotatebox{180}{\includegraphics{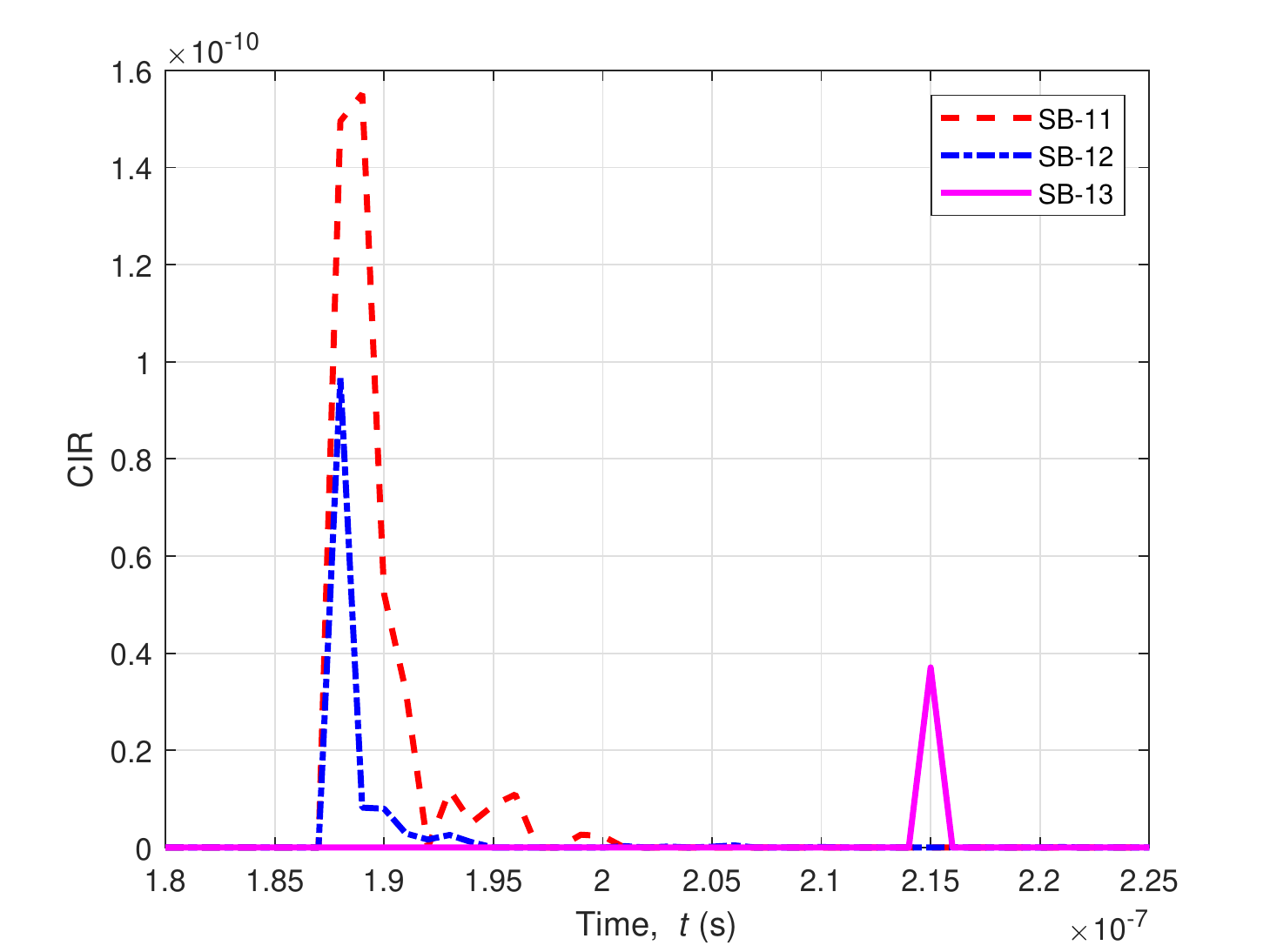}}}
\caption{CIR of the SB-11, SB-12 and SB-13 components in the geometric street angle scattering model.}
\vspace{-0cm}
\end{figure}\end{center}

\setlength{\unitlength}{1mm}\label{4}
\begin{center}\begin{figure}[h]
\scalebox{-0.59}{\rotatebox{180}{\includegraphics{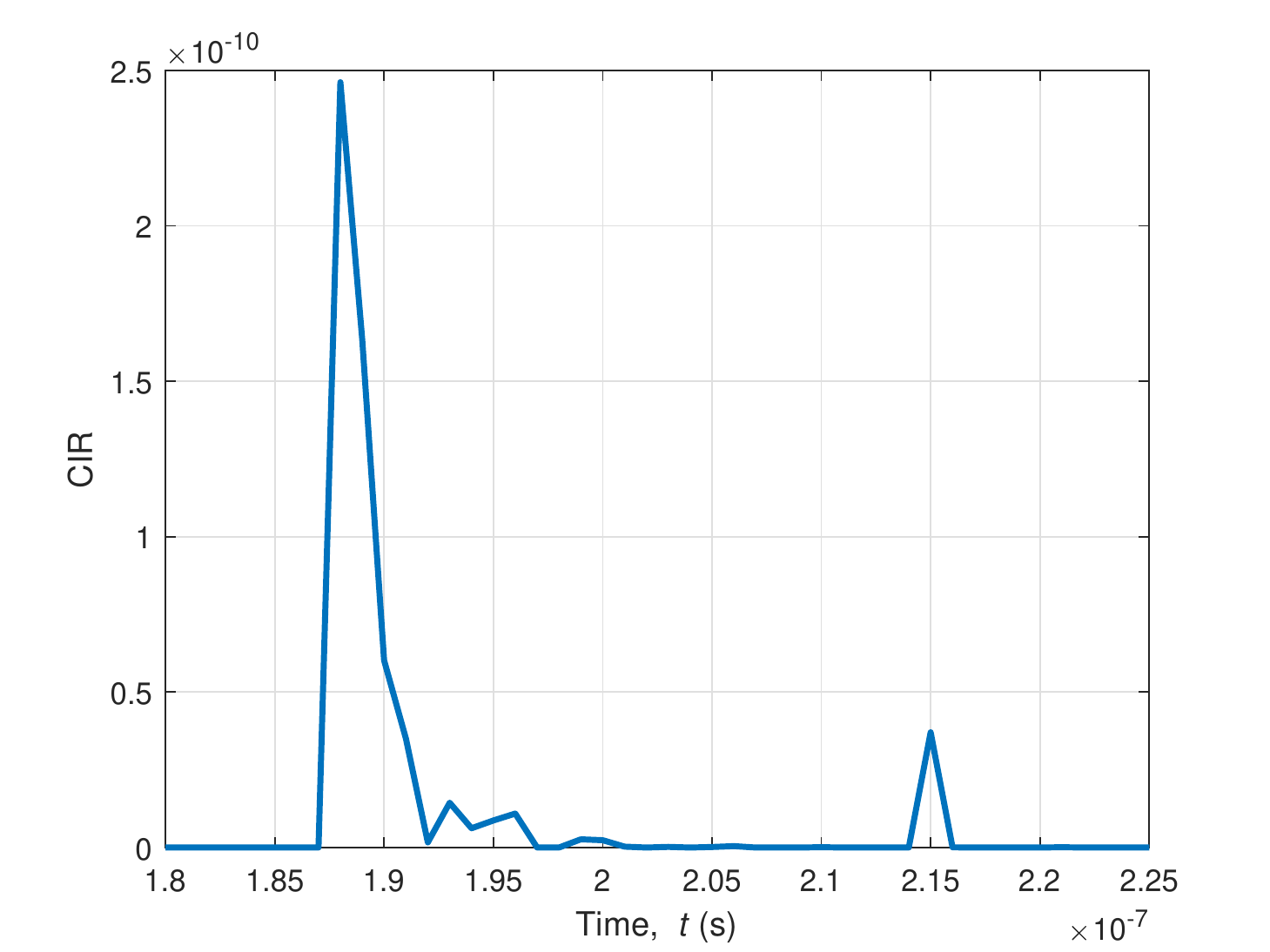}}}
\caption{Total CIR from the  geometric corner scattering model.}
\vspace{-3cm}
\end{figure}\end{center}

\setlength{\unitlength}{1mm}\label{5}
\begin{center}\begin{figure}[H]
\scalebox{-0.6}{\rotatebox{180}{\includegraphics{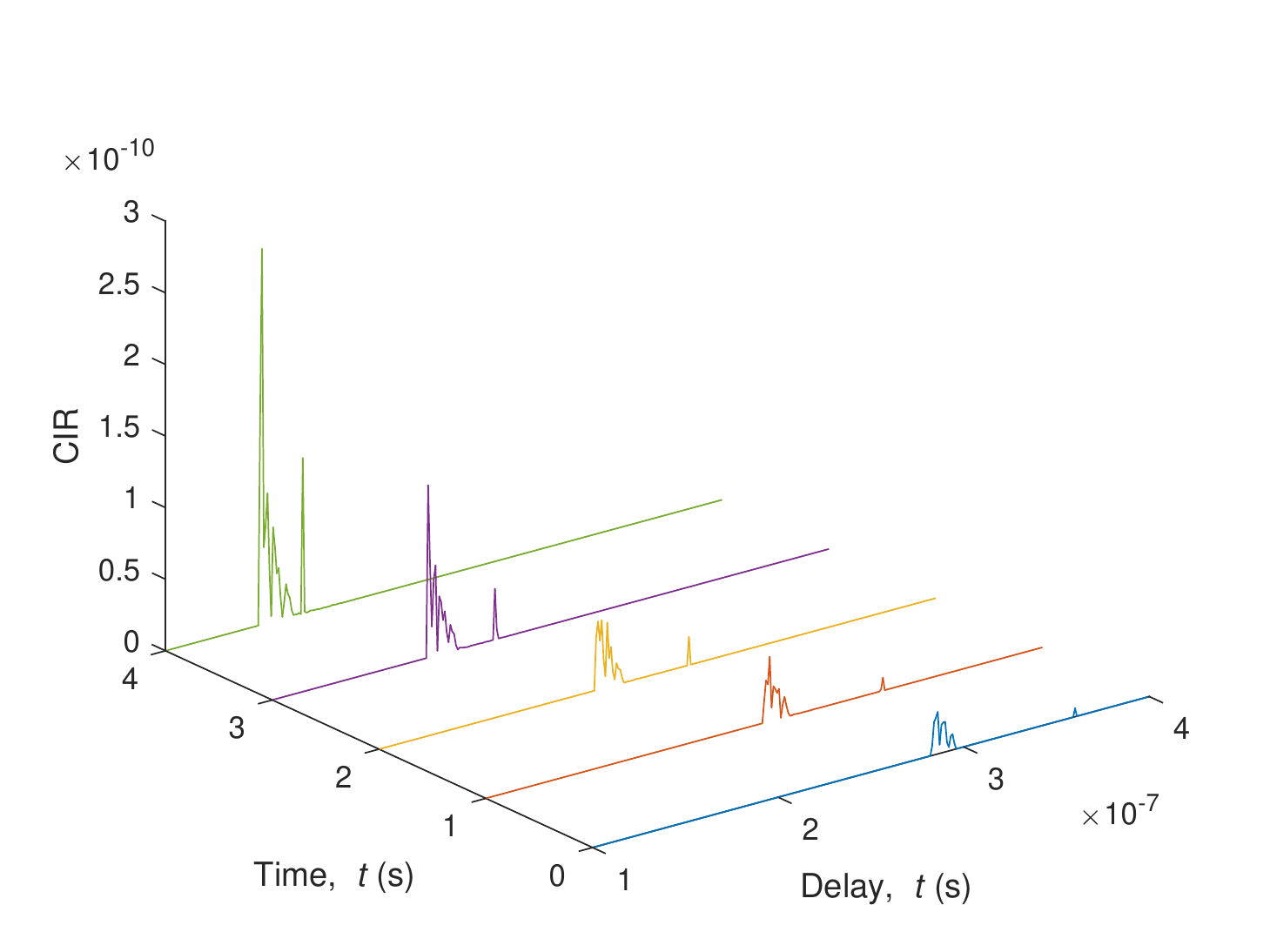}}}
\caption{CIR from SB component (the Tx and Rx are moving, $v_{\mathrm{Tx}}=25.2 \mathrm{km} / \mathrm{h}$, $v_{\mathrm{Rx}}=25.2 \mathrm{km} / \mathrm{h}$, $t=0-4 s$).}
\vspace{-2cm}
\end{figure}\end{center}

\setlength{\unitlength}{1mm}\label{6}
\setlength{\abovecaptionskip}{0.cm}
\begin{center}\begin{figure}[h]
\scalebox{-0.6}{\rotatebox{180}{\includegraphics{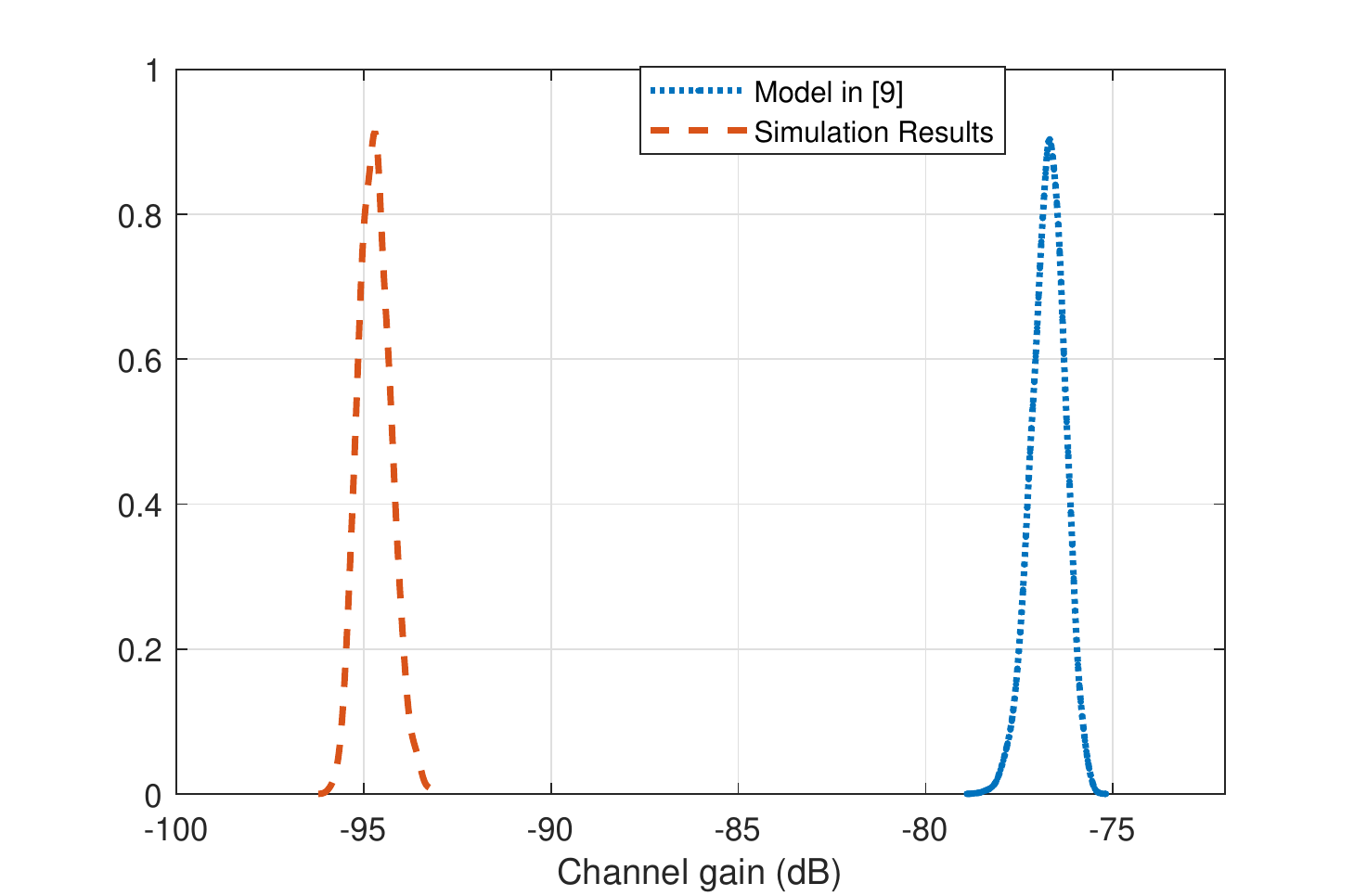}}}
\caption{Channel gain (dB) for SB components in geometric corner scattering model.}
\vspace{-1cm}
\end{figure}\end{center}

\section{CONCLUSION}

In this paper, a new non-stationary MIMO VVLC channel model has been proposed. The model considers both stationary and moving scatterers. Based on this model, the geometric characteristics of VVLC channel have been deduced, and the characteristics of VVLC channel have been studied deeply. The results show that the received light power increases and the time delay decreases as the distance decreases. We have researched the RMS delay spread and DC channel gain of the channel, we found that they were both Gaussian-distributed with the mean channel gain was -94.7dB. Finally, the simulation results have been compared with existing models and their distributions were found to be the same. The differences in results were also analyzed.

\section{ACKNOWLEDGEMENT}

This work was supported by the National Key R\&D Program of China under Grant 2018YFB1801101, the National Natural Science Foundation of China (NSFC) under Grant 61960206006, the High Level Innovation and Entrepreneurial Research Team Program in Jiangsu, the High Level Innovation and Entrepreneurial Talent Introduction Program in Jiangsu, the Research Fund of National Mobile Communications Research Laboratory, Southeast University, under Grant 2020B01, the Fundamental Research Funds for the Central Universities under Grant 2242019R30001, the Huawei Cooperation Project, the EU H2020 RISE TESTBED2 project under Grant 872172, the Taishan Scholar Program of Shandong Province, and Shandong Provincial Natural Science Foundation under Grant ZR2017MF012.

\setlength{\unitlength}{1mm}\label{7}
\begin{center}\begin{figure}[h]
\scalebox{-0.6}{\rotatebox{180}{\includegraphics{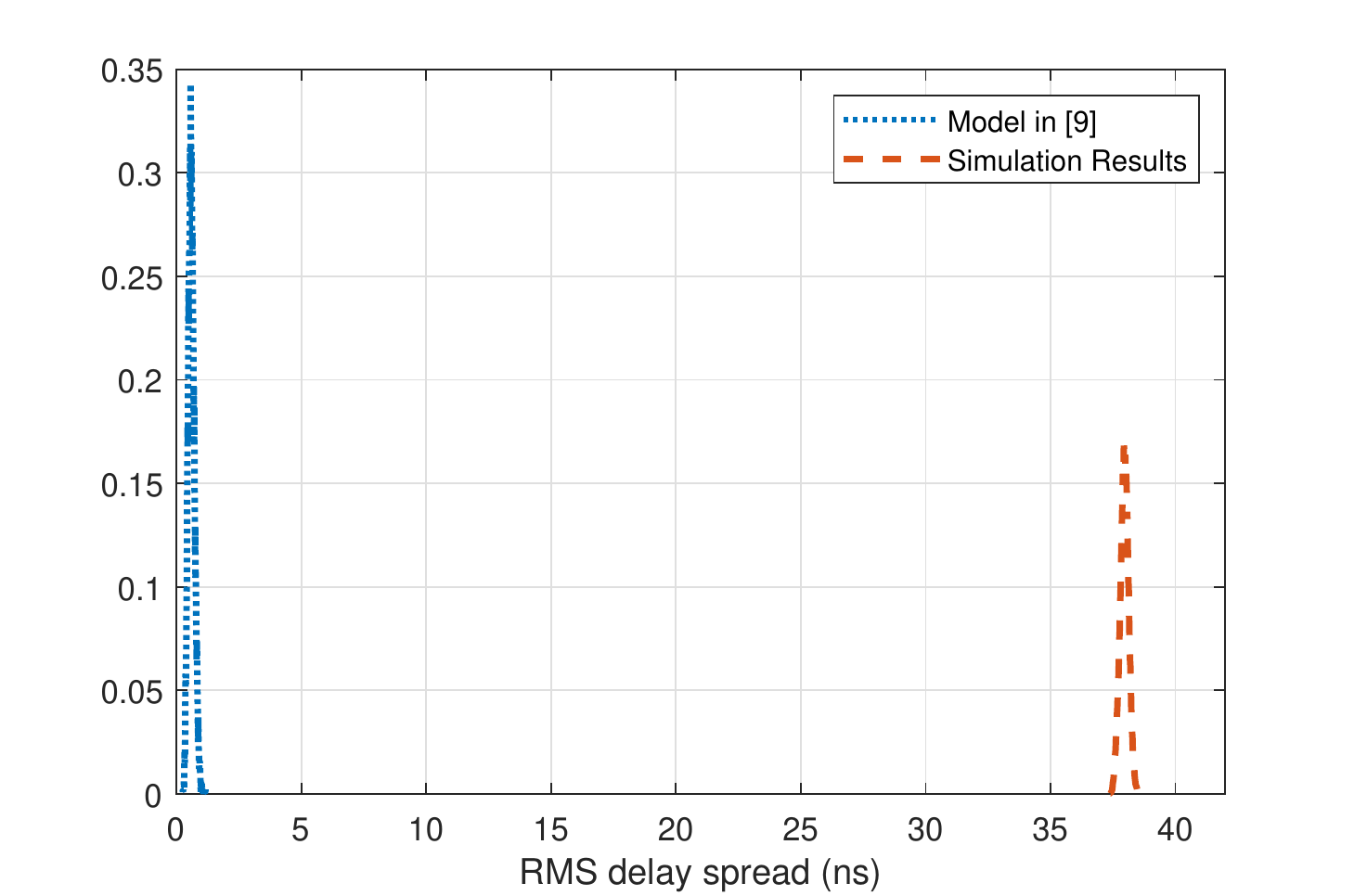}}}
\caption{RMS delay spread for SB components in geometric corner scattering model.}
\end{figure}\end{center}

\bibliographystyle{IEEEtrans}
\end{document}